\documentclass[aps,pre,twocolumn,groupedaddress]{revtex4}
\usepackage{graphicx,color,subfigure,amsmath}
\begin{document}
\title{Effect of chain stiffness on the competition between crystallization and glass-formation in model colloidal polymers}
\author{Hong T. Nguyen}
\author{Tyler B. Smith}
\author{Robert S. Hoy}
\email{rshoy@usf.edu}
\affiliation{Department of Physics, University of South Florida, Tampa, FL, 33620}
\author{Nikos Ch.\ Karayiannis}
\affiliation{Institute for Optoelectronics and Microsystems (ISOM) and ETSII, Polytechnic University of Madrid, Madrid, Spain}
\date{\today}

\begin{abstract}
We map out the solid-state morphologies formed by model soft-pearl-necklace polymers as a function of bending stiffness $k_b$ spanning the range from fully flexible to rodlike chains.
The ratio of Kuhn length to bead diameter ($l_K/r_0$) increases monotonically with increasing $k_b$ and yields a one-parameter model that relates chain shape to bulk morphology and yields insights into the packing of anisotropic particles.
In the flexible limit, monomers occupy the sites of close-packed crystallites while chains retain random-walk-like order.
In the rodlike limit, nematic chain ordering typical of lamellar precursors coexists with close-packing.
At intermediate values of bending stiffness the competition between random-walk-like and nematic chain ordering produces glass-formation; the range of $k_b$ over which this occurs increases with the thermal cooling rate $|\dot{T}|$ implemented in our molecular dynamics simulations.
Finally, values of $k_b$ between the glass-forming and rodlike ranges produce complex ordered phases such as close-packed spirals. 
Our results should prove useful for rational design of dense colloidal-polymer phases with desired morphologies.
\end{abstract}
\maketitle

\section{Introduction}

Over the last decade, advances in both experimental and computational techniques have contributed to rapid progress in our understanding of how complex, anisotropic particles pack.
Dense phases composed of such particles have been both synthesized in the laboratory and studied \textit{in silico}, and a remarkable ``zoology'' of packings has now been developed \cite{damasceno12}.
Phases as distinct as nematic crystals and amorphous glasses have been predicted, and are plausible candidates for the packings of natural anisotropic constituents ranging from sand grains to viruses.
One class of such complex systems is that of atoms/particles/molecules with highly anisotropic but \emph{fixed} shape \cite{damasceno12,donev04,donev04b,jiao09,torquato09,haji11a,haji11b,ni12}.
Another class consists of entities (e.g.\ molecules) whose shape is not only anisotropic but fluctuates over the observation time.
Many such systems exist in nature, prominent among them being the synthetic and biological polymers.
These macromolecules exhibit a wide variety of shapes and structural characteristics leading to distinct phase behaviors that depend on chain topology/connectivity, chemistry (e.g.\ monomer shape) and  bond and torsion angle stiffness.
The structural motifs formed by ``traditional'' polymers are diverse, and thus difficult to characterize precisely and to study in parametric fashion, especially at the atomistic level.

\textit{Colloidal} polymers (CPs) composed of chains of linked, macroscopic monomers \cite{miskin13,miskin14,zou09,sacanna10,brown2012,blaaderen2012,reichhardt2011} offer a promising alternative for studies of the packing ability and phase behavior of systems formed by chain molecules.
For example, one controllable parameter in CPs is the ratio $l_K/r_0$ of the Kuhn length $l_K$, i.e.\ the length over which chains appear rodlike, to the monomer diameter $r_0$.
Controlling this ratio can naturally be expected to profoundly affect the morphologies formed by CPs.
However, experimental study of CPs remains in its infancy.
Only a few systems have been synthesized, and the factors affecting their packing at both the monomer and chain scales remain poorly explored.
Thus there is a great need to characterize how parameters such as $l_K/r_0$ affect the morphologies of dense phases.

The main goal of this study is to examine the role chain topology and chain flexibility (or equivalently stiffness) plays in controlling solidification, i.e.\ crystallization vs.\ glass-formation, in model colloidal polymers.
We perform extensive molecular dynamics (MD) simulations to determine how chains of varied bending stiffness $k_b$, spanning the entire range from the flexible to the rodlike limit, solidify during cooling, and identify the factors controlling crystallization, glass-formation, and the competition thereof.
Through these simulations we are thus able to identify how characteristic features of the solidification process vary with $k_b$, and thus establish the effect of chain stiffness on the morphologies formed during cooling from the disordered liquid phase.

Our results indicate that both isotropic and nematic systems crystallize, but into distinctly different ordered phases.
In the flexible limit ($l_K \simeq r_0$), monomers occupy the sites of close-packed (FCC or HCP) crystallites, but chains possess random-walk-like structure.
In the rodlike limit ($l_K \gg r_0$), monomers again close-pack, but chains form nematic domains with a single, well-defined orientation.
For intermediate bending stiffness systems possess chain-scale ordering intermediate between isotropic and nematic, and form solids that are highly disordered at the monomer level, i.e.\ are good glass-formers.
Neither intermediate values of $l_K/r_0$ nor intermediate-scale nematic order seem to be compatible with close-packing; glass formation is associated with the resulting geometric frustration.
We also report novel spiral and multidirectional-nematic phases that coexist with locally close-packed order.
The broad array of morphologies predicted by our single-parameter model suggests that varying the associated, aspect-ratio-like quantity $l_K/r_0$ in ``pearl-necklace-like'' colloidal polymers could be used to control both their global (chain-scale) and local (monomer-scale) structure.

\section{Model and Methods}

\subsection{Interaction potential and MD simulation protocol} 

Our simulations employ the semiflexible version of the soft-colloidal-polymer model described at length in Ref.\ \cite{hoy13}.
It is comparable to the Kremer-Grest bead-spring model \cite{kremer90}, but possesses crystalline ground states since the equilibrium bond length $\ell_0$ is commensurate with the monomer diameter $r_0$, i.e.\ the condition $\ell_0=r_0$ is amenable to formation of close-packed structures.  
All monomers have mass $m$ and interact via the truncated and shifted Lennard-Jones potential
\begin{equation}
U_{LJ} = 4\epsilon\left[\left(\displaystyle\frac{\sigma}{r}\right)^{12} - \left(\displaystyle\frac{\sigma}{r}\right)^{6} - \left(\displaystyle\frac{\sigma}{r_c}\right)^{12} + \left(\displaystyle\frac{\sigma}{r_c}\right)^{6}\right],
\label{eq:LJpot}
\end{equation}
where $\epsilon$ is the intermonomer binding energy and $r_c$ is the cutoff radius.
Attractive van der Waals interactions are included by setting $r_c = 2^{7/6}$ (in LJ units).
The MD timestep used here is $\delta t = \tau/200$, where $\tau$ is the Lennard-Jones time unit $\sqrt{ma^2/\epsilon}$.

Bonds between adjacent beads along the chain backbone are modeled using the harmonic potential
\begin{equation}
U_c(r) = \displaystyle\frac{k_c}{2}\left(r-a\right)^2,
\label{eq:Ubond}
\end{equation}
where $a$ is the monomer diameter and $k_c = 600\epsilon/a^2$ is the bond stiffness.
To produce polymer chains with $\ell_0 = r_0$, we set $\sigma = 2^{-1/6}a$.
For this stiffness, the energy barrier for chain crossing is at least $50k_BT$ over the whole temperature range considered herein.
Bending stiffness is included using the standard potential \cite{auhl03}
\begin{equation}
U_b(\theta) = k_b(1 - cos(\theta)),
\label{eq:Ubend}
\end{equation}
where $cos(\theta_i) = (\vec{b}_i\cdot\vec{b}_{i+1})/(\|\vec{b}_{i}\|\|\vec{b}_{i+1}\|)$ and the bond vector $\vec{b}_i = \vec{r}_{i+1}-\vec{r}_i$.
We study systems with $0 \leq k_b \leq 12.5\epsilon$.

All systems are composed of $N_{ch}=500$ chains.
Periodic boundaries are applied along all three directions of cubic simulation cells.
In this study we focus on polymers with chain lengths $N = 13,\ 25, \textrm{and}\ 50$ and compare them to results for monomers.
This range of $N$ is typical of colloidal polymers synthesized and studied experimentally\ \cite{zou09,sacanna10,blaaderen2012,brown2012,miskin14}.
Initial systems are generated by placing randomly oriented random-walk-like coils within these cells.
Systems are then thoroughly equilibrated at temperatures (monomer number densities) $k_BT/\epsilon = 1.2$ ($\rho = 1.0a^{-3})$ for $k_b < 7\epsilon$, $k_BT/\epsilon = 1.4$ ($\rho = 0.9a^{-3}$) for $7\epsilon \leq k_b < 10\epsilon$, and $k_BT/\epsilon = 1.6$ ($\rho = 0.8a^{-3}$) for $k_b \geq 10\epsilon$, i.e.\ well above the $k_b$-dependent solidification temperatures reported below.
To avoid any artifacts arising from insufficient equilibration, we monitor the chain statistics $\left<R^{2}(n)\right>$ and check for convergence at all chemical distances $n$ \cite{auhl03}.
Systems are then further equilibrated at constant (zero) pressure, and then cooled to $T=0$ (also at zero pressure) at a rate $|\dot{T}|$.
Pressure is controlled using a Nose-Hoover barostat.
To examine the dependence of solid-state ordering upon the cooling (quench) rate, we consider  $|\dot{T}|$ ranging from $10^{-6}/\tau$ to $10^{-4}/\tau$.
All MD simulations are performed using LAMMPS \cite{plimpton95}.

\subsection{Measures of monomer- and chain-scale order}

During cooling we monitor several metrics quantifying local (monomer-scale) and global (chain-scale and above) structure.
Since $U_{LJ}$ is attractive and we perform simulations at zero pressure, all systems densify with decreasing $T$.
We report the monomer densities $\rho(T)$ in terms of the packing fraction $\phi = \pi\rho/6$.
Since $U_{LJ}$ is a ``soft'' potential and solidification occurs at rather high $T$, and to aid comparison with results for other models including athermal (hard-sphere) systems (e.g.\ \cite{karayiannis09,karayiannis10,karayiannis13}), we also report values of the \textit{effective} (thermalized \cite{zhang08}) packing fraction at solidification: $\phi_s^{eff} = \pi\rho_s^{eff}/6$, where $\rho_s^{eff} = \rho (r_s^{eff})^{3}$, and the effective monomer radius $r_s^{eff}$ is the smallest real solution to \cite{prefac}
\begin{equation}
U_{LJ}(r_s^{eff}) - U_{LJ}(r_0) = 1.1k_B T_s.
\label{eq:phiseff}
\end{equation}

Local structure at the monomer level is identified through the characteristic crystallographic element (CCE) norms \cite{cce09,karayiannis09,karayiannis10,karayiannis13}.
The CCE-based analysis employs highly discriminating descriptors that quantify the orientational and radial similarity of a given monomer's local environment to that of various ordered structures.
CCE norms are built around the defining set of crystallographic elements and the subset of distinct elements of the corresponding point symmetry groups that uniquely characterize the reference crystal structure. For example, the FCC crystal symmetry is mapped onto a set of four three-fold axes (roto-inversions of $2\pi/3$), while the HCP is mapped onto a single six-fold symmetry axis (roto-inversion of $\pi/3$). A scan in the azimuthal and polar angles identifies the set of axes that minimize the CCE norm of a reference site (atom or particle) with respect to a given crystal structure $\emph{X}$.  Details on the underlying mathematical formulae and the algorithmic implementation can be found in Ref.\ \cite{cce09}. Once the CCE norm ($\epsilon_{i}^{X}$) is calculated for each site $\emph{i}$ an order parameter $s^{X}$ can be calculated, which corresponds to the fraction of sites with CCE norms below a pre-set threshold value ($\epsilon_{i}^{X}\le\epsilon^{\rm thres}$). Here, CCE norms are calculated with respect to the face centered cubic (FCC) and hexagonal close packed (HCP) crystals, and the fivefold local symmetry.

Multiple measures are used to characterize nematic order.
The polymers' persistence length \cite{degennes79}
\begin{equation}
l_p = \sum_{i=0}^{N-2} \vec{b}_i\cdot\vec{b}_{i+1},
\label{eq:perslength}
\end{equation}
measures how single-chain conformations change with $T$, i.e.\ chain folding/unfolding.
Average nematic order at the \textit{chain} level is characterized via the method employed in Ref.\ \cite{degennes79}:
alignment of chains can be characterized by the largest eigenvalue $S_{g}$  of the tensor $Q_{\alpha \beta}$ which is defined by
			\begin{equation}
				Q_{\alpha\beta} = \biggl\langle \frac{3}{2}\hat{u}_{j\alpha} \hat{u}_{j\beta}
								- \frac{1}{2} \delta_{\alpha\beta} \biggl\rangle,
				\label{eq:1}
			\end{equation}
where $\hat{u}_{j}$ is the end-to-end unit vector of chain $j$, $\delta$ is the Kronecker delta and $\alpha,\beta$ denote the Cartesian directions $x,y,z$, and the average is taken over all chains in the system.
By construction, $S_{g}$ = 1 signifies perfect alignment of all chains. In contrast, $S_{g}$ = 0 means random orientation.

Nematic order at the \textit{bond} level is characterized via the method employed in Ref.\ \cite{luo11}: tensor order $S$ is given by
			\begin{equation}
				S=\sqrt {\frac{3}{2} Tr( q^2)},
				~q_{\alpha\beta}= \biggl\langle \hat{b}_{\alpha}\hat{b}_{\beta} -\frac{1}{3} \delta_{\alpha\beta}  \biggl\rangle .
				\label{eq:2}
			\end{equation}
Here, Tr is the trace operator, $\bigl\langle \cdots \bigr\rangle$ denotes the average over all normalized bond vectors ${\bf b}$ in each sub-cell, and $\hat{b}_{\alpha}$ and $\hat{b}_{\beta}$ are Cartesian components of ${\bf b}$.
In Equation \ref{eq:2}, $S = 1$ corresponds to perfect alignment of bonds in a given subcell and S = 0 corresponds to random bond orientation within that subcell.
In order to get a single number for the average bond orientation in the system, we average S over all subcells in the simulation \cite{footSsubcell}.  While the average tensor order defined in this way depends on the size of the subcells used, we have
tested different grid sizes and found that the results presented below are qualitatively unaffected by small changes when the subcell size is 2-3 monomer diameters.
Another measure of nematic order at the bond level is provided by the bond-orientational correlation function
\begin{equation}
F_{bb}(r) = \left<\left|\vec{b_i}(\vec{r}_{0})\cdot\vec{b_j}(\vec{r}_0 + \vec{r})\right|\right> - \displaystyle\frac{1}{2},
\label{eq:bonddircorr}
\end{equation}
a sensitive measure of long-distance nematic order that is positive when bond vectors separated by a distance $r$ are correlated, and zero when they are uncorrelated.
Here $\vec{b}_i = \vec{r}_{i+1}-\vec{r}_i$ as defined above, $r_0 = (\vec{r}_{i+1}+\vec{r}_i)/2$ indicates the midpoint of this bond, $\vec{r}_0 + \vec{r} = (\vec{r}_{j+1}+\vec{r}_j)/2$ ndicates the midpoint of the $j_{th}$ bond, and the brackets denote averages over all $j > i$.

\subsection{Distinction from previous modeling efforts}

In the original Kremer-Grest bead-spring model \cite{kremer90} and subsequent modifications \cite{abrams01,mackura13}, the equilibrium backbone bond length $\ell_0$ is different from the equilibrium non-bonded separation $r_0$; these competing length scales were shown in Refs.\ \cite{bennemann98,abrams01,mackura13} to suppress crystallization \cite{KGliqcryst}.
Our model sets $\ell_o = r_o$  and therefore possesses a simple (close-packed) crystalline ground state.
However, it also displays glass-formation under thermal quenches at sufficiently large $|\dot{T}|$ \cite{hoy13}, and is therefore well suited to studies of competing crystallization and glass transitions.

Another class of widely studied polymer model from which ours is crucially different is tangent hard spheres (THS).
THS polymers may be either flexible (freely-jointed) \cite{vega01,vega02,karayiannis08,karayiannis10,karayiannis13} or semiflexible \cite{fynewever98}.
They are \textit{athermal} since monomers interact via purely repulsive (hard-sphere) pair potentials.
Further, they usually fix $l_o =  r_o$ through holonomic constraints rather than the relatively ``soft'' harmonic potential (Eq.\ \ref{eq:Ubond}) used herein.  
Relaxation of the $\ell_0=r_0$ condition \cite{ni13} has been shown to profoundly affect THS crystallizability, in particular by speeding its dynamics \cite{ni13}, or by changing the ground state's order by imposing $\ell_0 \neq r_0$ \cite{nck15}.
However, the present model should not be confused with these; here $\ell_0 = r_0$ applies only in a thermodynamically averaged sense, since bond length can fluctuate significantly due to thermal excitations or to structural disorder at $T=0$ \cite{hoy13}.

The ``partially flexible'' (PF) model \cite{oyarzun13,westen13} is comparable to semiflexible THS; part of each chain is maintained in a linear configuration (i.e.\ a rod), and the remainder is completely flexible (freely-jointed).
Chain rigidity in the PF model is characterized by the fraction of monomers $f_r$ in the rigid segment, e.g. monomers are arranged into rigid linear rods for $f_r=1$ \cite{vega01,vega02}, whereas $f_r = 0$ corresponds to flexible THS.
In contrast, chain rigidity in our model is constant along chains and is specified by the strength $k_b$ of angular interactions between three consecutive monomers (Eq. \ref{eq:Ubend}).

MD simulations similar to those employed here have long been used to examine ordered structures formed during quenching; see e.g.\ Refs.\ \cite{meyer01,meyer02,vettorel07,sommer10,luo13}.
Vettorel et.\ al.\ \cite{vettorel07} employed very similar simulations to study the ``phase diagram'' of coarse-grained PVA; Figure 1 of Ref.\ \cite{vettorel07} is very similar to our Figure \ref{fig:PD}.
A key difference is that the CG-PVA model employs fixed interaction potentials (producing relatively stiff chains), whereas here we vary chain stiffness over the entire range from fully-flexible to rodlike via the parameter $k_b$.
Finally, a very recent study \cite{kumar13} employed a KG-like model with variable chain stiffness to study the influence of stiffness on chain packing and fragility \cite{ediger00} in glass-forming systems; investigations of the latter using our crystallizable model are under way and will be reported in a forthcoming publication.

\section{Results}

\subsection{Preparation-protocol dependence}

\begin{figure}[htbp]
\centering
\includegraphics[width=2.5in]{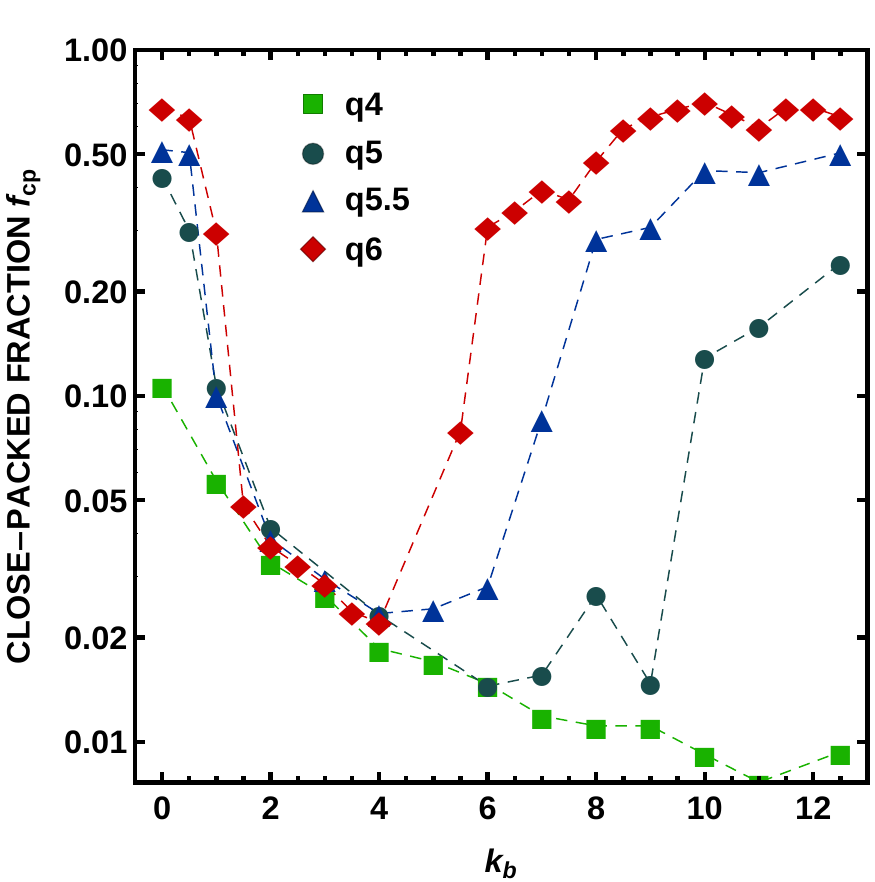}
\caption{Quench rate dependence of $f_{cp}$ values in the $T=0$ end states of cooling runs for all $N=25$ systems.  In the legend, ``q\textit{x}'' indicates $|\dot{T}| = 10^{-x}/\tau$.}
\label{fig:QRdep}
\end{figure}

Real colloidal polymers often glass-form \cite{zou09}.
This may be due either to geometrical factors such as incommensurability of their bond length and bead diameter, or to details of the protocol with which they are prepared.
While our model is commensurable and possesses crystalline ground states, MD simulations show that the morphologies formed during solidification are strongly preparation-protocol dependent.
Figure \ref{fig:QRdep} shows results for the close-packed monomer fraction $f_{cp}$, which is the summation of the HCP- and FCC-like site fractions, in the ($T=0$) end states of cooling simulations performed over a range of rates $10^{-4}/\tau \leq |\dot{T}| \leq 10^{-6}/\tau$.
For $|\dot{T}| = 10^{-6}/\tau$, $f_{cp}$ varies very strongly with $k_b$ as described and analyzed below.
Values of $f_{cp}$ at fixed $k_b$ vary differently with $|\dot{T}|$ for different $k_b$;
while all decrease with increasing $|\dot{T}|$, the strength of this decrease is highly $k_b$-dependent.

These results illustrate two key features of our model and simulation method: \textbf{(i)} the "critical" cooling rates below which we obtain a large $f_{cp}$ vary significantly with $k_b$; \textbf{(ii)} for all $k_b$, the variation in the obtained morphology is significant over our achievable range of cooling rates.
The cooling rate $|\dot{T}|$ also exerts profound influence on the $k_b$-dependence of the obtained morphologies reported below in Figure \ref{fig:PD}.  
In particular, with increasing $|\dot{T}|$, the range of $k_b$ over which systems glass-form expands outwards in both directions.  
{Since similar considerations are expected to apply for real CPs, it is important to identify critical cooling rates (or their equivalents in athermal systems) below which experimental systems will form desired ordered morphologies.

The lowest $|\dot{T}|$ feasible given current computational power is $10^{-6}$.
This rate produces a remarkably strong $k_b$ dependence (maximal values of $f_{cp}$ are larger than minimal values by nearly two orders of magnitude), and as will be described below, tremendous diversity in the morphologies formed upon solidification.
We therefore focus on results from $|\dot{T}|=10^{-6}/\tau$ quenches throughout the remainder of the paper.

\subsection{Solidifcation densities and temperatures}

The simplest structural metric characterizing the phase behavior and in particular the competition between crystallization and glass formation is the temperature dependence of the packing fraction $\phi = \pi\rho/6$. 
Figure \ref{fig:phi} shows results for $\phi(T)$ for all systems.
At high $T$, systems show a linear increase in $\phi$ as $T$ decreases, i.e. they densify with a constant thermal expansion coefficient.
For most systems this increase persists until the onset of solidification, but for the stiffest systems it is interrupted by a density increase corresponding to the isotropic-melt$\to$nematic-melt transition (discussed further below.)
At the solidification temperature $T=T_s$,  $\phi$ increases more rapidly for crystal-forming systems; this increase is a sharp, first-order-like jump for the lowest and highest values of $k_b$ and a more gradual concave-up increase for intermediate stiffness.
Glass-forming systems show a concave-down increase. 
Values of $T_s$ (reported below in Fig.\ \ref{fig:PD}) are identified with first-order like jumps or the points of inflection of $\phi(T)$ ($T_{cryst}$) or  the intersection of linear fits to the high-$T$ and low-$T$ regimes of $\phi(T)$ ($T_g$).
Both $T_{cryst}$ and $T_g$ increase monotonically with increasing $k_b \lesssim 2\epsilon$, as is expected for traditional polymer chains of increasing stiffness \cite{dudowicz05,kunal08,schnell11}.
Finally, after solidification, all systems again show a linear increase in $\phi$ as $T$ continues to decrease towards zero.

\begin{figure}[htbp]
\centering
\includegraphics[width=3in]{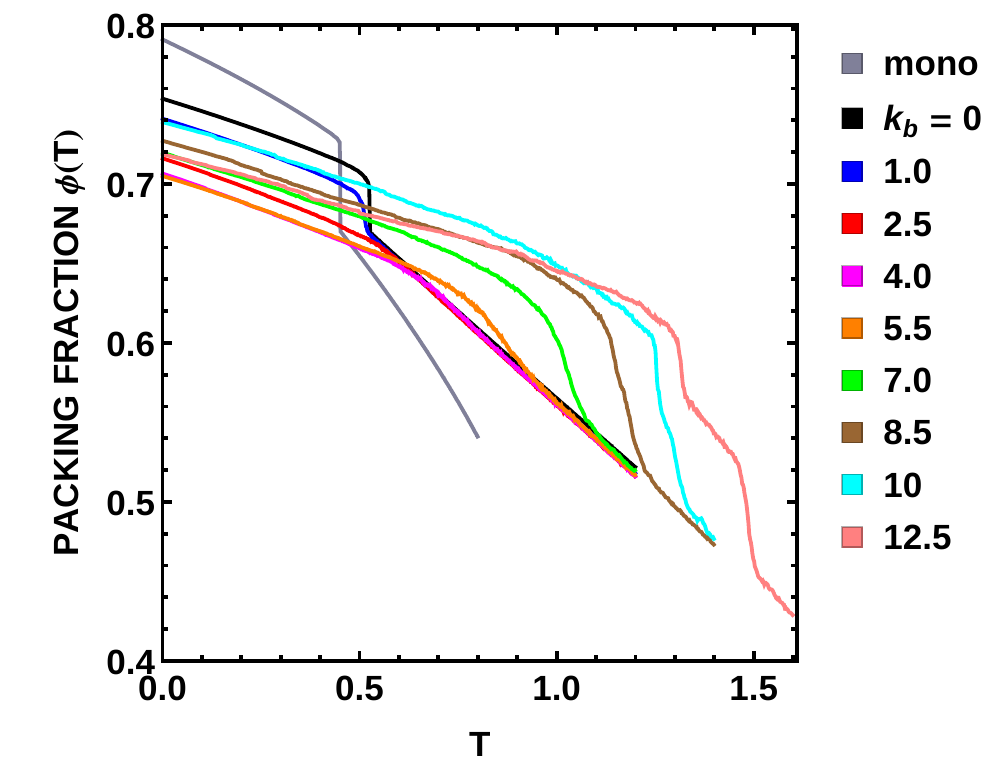}
\caption{Thermodynamic signatures of crystallization (or the absence of it): packing fraction $\phi(T)$ for selected $N= 25$ systems.  Data for monomers ($N=1$) are included for comparison.}
\label{fig:phi}
\end{figure}

It is interesting to examine the $k_b$-dependence of system densities at solidification.
Values of $\phi_s(k_b) = \phi(T=T_s(k_b))$ are reported in Table \ref{tab:phisolid}.
They decrease nearly monotonically with $k_b$; flexible systems crystallize at $\phi_s \simeq .68$, glassformers solidify in the range $0.6 < \phi_s < 0.67$, and  for $k_b \gtrsim 7.5\epsilon$ all systems crystallize at $\phi_s \simeq .58$.   
Values of $\phi_s^{eff}$ also decrease nearly monotonically with increasing $k_b$, from very slightly above the hard-sphere jamming fraction $\phi_J = .637$ \cite{torquato00} to about $0.52$ for rodlike chains.
While a decrease is expected since rod-like particles generally jam or crystallize at lower $\phi$ than their spherical counterparts \cite{kyrylyuk11}, we are not aware of any previous studies that systematically examined solidification density as a function of chain stiffness / aspect ratio in model polymers.
Further, we will show below that the crossover from higher to lower values of $\phi_s^{eff}$ that occurs at $k_b \sim 6$ corresponds to the onset of local nematic order (cf.\ Fig.\ \ref{fig:Fbb}.)

\begin{table}[htbp]
\caption{Values of $\phi$ upon solidification for all $N=25$ systems: $\phi_s(k_b) = \phi(T=T_s(k_b))$ and $\phi_s^{eff}(k_b)$ (Eq.\ \ref{eq:phiseff}) for the simulations and values of $T_s$ reported in Figure \ref{fig:PD} (as well as similar simulations of several additional $k_b$.)  For comparison, monomers have $\phi_s = .6876$ and $\phi_s^{eff} = 0.650$.
We estimate ``error bars'' on all measured quantities are of order $1\%$.}
\begin{ruledtabular}
\begin{tabular}{lccccc}
$k_b/\epsilon$ & $\phi_s$  & $\phi_s^{eff}$ & $k_b/\epsilon$ & $\phi_s$ & $\phi_s^{eff}$ \\
0    & 0.683 & 0.643 & 6.5	& 0.590 & 0.535 \\
0.5 &  0.683 & 0.643 & 7	& 0.592 & 0.535 \\
1   & 0.679 & 0.641 & 7.5	& 0.582 & 0.522 \\
1.5 & 0.681 & 0.645 & 8	& 0.581 & 0.520 \\
2	& 0.673 & 0.636 & 8.5 & 0.570 & 0.508 \\
2.5	& 0.666 & 0.627 & 9    & 0.571 & 0.518 \\
3	& 0.657 & 0.616 & 9.5 & 0.586 & 0.520 \\
3.5	& 0.651 & 0.609 & 10	& 0.580 & 0.514 \\
4	& 0.648 & 0.605 & 10.5  & 0.580 & 0.514 \\
4.5	& 0.643 & 0.599 & 11	& 0.578 & 0.512 \\
5	& 0.642 & 0.598 & 11.5 & 0.583 & 0.515 \\
5.5	& 0.597 & 0.544 & 12	& 0.580 & 0.512 \\
6	& 0.606 & 0.552 & 12.5 & 0.584 & 0.526 \\
\end{tabular}
\end{ruledtabular}
\label{tab:phisolid}
\end{table}

Studies of jamming in thermalized colloidal systems \cite{zhang08} have similarly found $\phi_s^{eff} \simeq \phi_J$, and that crystallization sets in as $\phi$ increases beyond $\phi_J$. 
We therefore conclude that results expressed in terms of $\phi_s^{eff}$ should be valuable in mapping our results to those obtained in experiments as well as simulations where particles typically interact via stiffer potentials.

\subsection{Local ordering from monomeric to segment scales}

Next we show that the trends in $\phi(T)$ are closely matched by corresponding ones obtained from different descriptors of local  order.
Figure \ref{fig:fcpf5f} shows the $T$-dependent fractions $f_{cp}(T)$ and $f_{5f}(T)$ of monomers with (a) close-packed order (FCC or HCP similarity) and (b) fivefold local symmetry, as quantified by CCE analysis \cite{cce09}.
Systems exhibiting sharp jumps in $\phi(T)$ also show sharp jumps in $f_{cp}(T)$.
For very flexible and very stiff chains these jumps are reminiscent of first-order transitions and occur at $T = T_{cryst}$.
In contrast, glassy systems (e.g.\ $k_b = 4\epsilon$) show a more gradual and much weaker increase, and ultimately exhibit far lower ultimate values of $f_{cp}$.
Results for $f_{5f}(T)$ display opposite trends.
For glassforming systems, $f_{5f}$ increases continuously with decreasing $T$, as expected in a densifying glassformer \cite{frank52}.
In the limit of fully flexible chains ($k_b = 0$), the sharp increase of ordered sites is accompanied by a sharp decline in the fivefold population.
In contrast, for chains that order nematically, $f_{5f}$ remains low at all $T$.
This is expected since the hexagonal order in planes perpendicular to the nematic director field \cite{meyer01,meyer02,Vettorel2007} suppresses fivefold order.

\begin{figure}[htbp]
\centering
\includegraphics[width=2.6in]{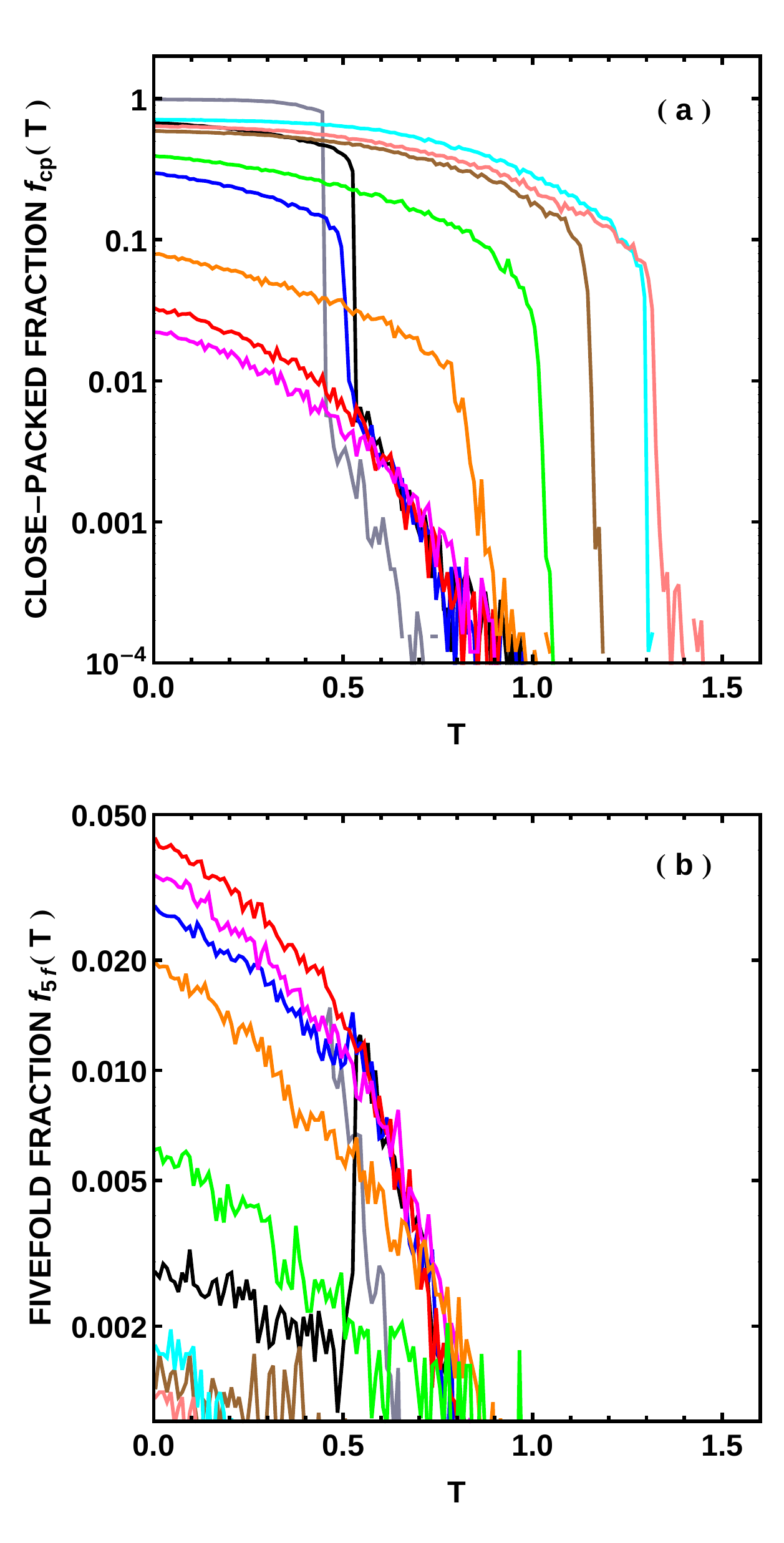}
\vspace{-10pt}
\caption{Thermodynamic signatures of crystallization  (or the absence of it) for selected $N=25$ systems: (a) Close-packed fraction $f_{cp}(T)$, (b) fivefold fraction $f_{5f}(T)$.  Colors are the same as in Fig.\ \ref{fig:phi}.}
\label{fig:fcpf5f}
\end{figure}

Simple bead-spring models like ours possess two essentially ``polymeric'' features controlling the crystallization-vs-glass-formation competition: topological chain connectivity / uncrossability, and angular stiffness.
Here the role of chain topology is indicated by contrasting results for polymers to data for monomers presented in Figures \ref{fig:phi}-\ref{fig:fcpf5f}.
The monomeric Lennard-Jones system is well-known as an excellent crystal-former \cite{trudu06,wang07}.
Monomers exhibit sharper transitions in $\phi$ and $f_{cp}$, and ultimately reach higher values of both order metrics at $T=0$.  Furthermore, they exhibit significantly lower $f_{5f}$ as the resulting ordered morphology is an almost perfect FCC crystal.
These differences arise because chain connectivity reduces both the critical rates for crystal nucleation and growth and the entropy of close-packed crystallites.
Simulations with shorter and longer chains indicate that the aforementioned trends strengthen with increasing $N$, especially once $N$ increases beyond the onset of chain entanglement \cite{hoy09}, and especially for stiffer chains.

Angular stiffness effects on local (dis)ordering propensity can be readily examined though analyzing distributions of bond and torsion angles (respectively $\theta$ and $\psi$).
Figure \ref{fig:angledist} shows the probability distributions $P(\theta)$ and $P(\psi)$ in the $T=0$ end states of of cooling runs for selected $k_b$.
For flexible chains, peaks are in $P(\theta)$ are observed at $0^{\circ}$, $60^{\circ}$, $90^{\circ}$, and $120^{\circ}$, characteristic of a stack-faulted close-packed structure \cite{hoy13} and similar to that observed in crystallized athermal polymers \cite{karayiannis10}.
The peaks characteristic of crystalline order decrease in intensity as $k_b$ increases, and vanish by $k_b=2.5\epsilon$, being replaced by a single broad peak at large $\theta$ (e.g.\ as shown for $k_b=4\epsilon$).
We claim that locally amorphous order arises because this large, broad peak is incompatible with close-packed ordering, e.g. chains are too stiff (flexible) to form the characteristic $120^{\circ}$ ($0^{\circ}$) angles with high probability at temperatures near solidification.
In other words, chains in glassforming systems are too stiff to collapse into random-walk configurations and form close-packed (RWCP) crystals, but not stiff enough to form the extended nematic domains essential for crystallization into the nematic close-packed (NCP) phase.
Thus crystallization is hindered and systems remain amorphous during solidification.
As stiffness continues to increase, however, the abovementioned broad peak is replaced by a sharp peak at $\theta=0^{\circ}$ (illustrated here in $P(\theta)$ for $k_b = 12.5\epsilon$), indicating increasingly rodlike configurations that can efficiently close-pack and form NCP crystallites.

\begin{figure}[h!]
\centering
\includegraphics[width=2.6in]{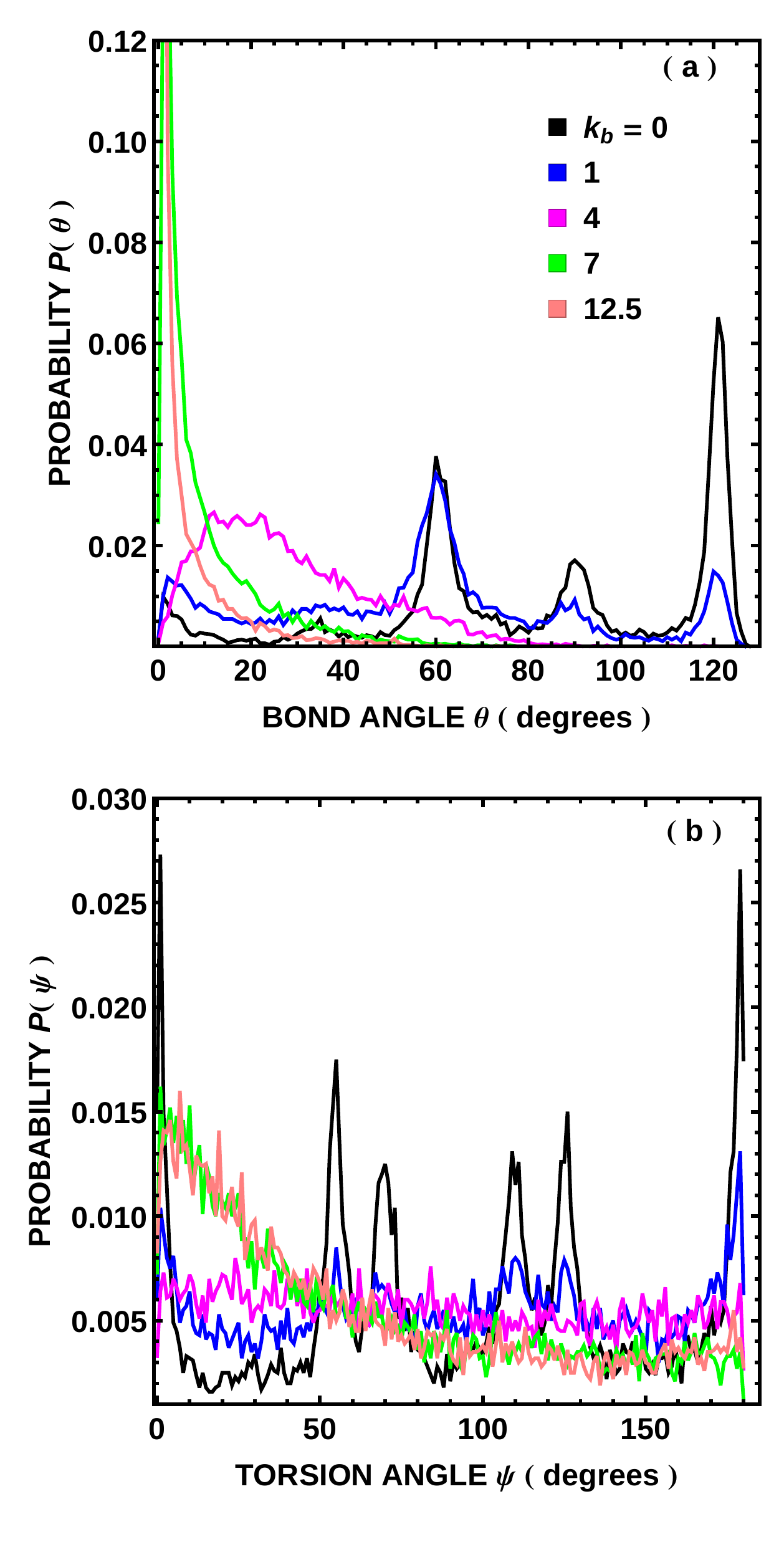}
\vspace{-10pt}
\caption{Probability distributions of (a) bond angles and (b) torsion angles at $T=0$  for selected $N=25$ systems:.  The $0^{\circ}$ ``origins'' respectively correspond to straight trimers and trans conformers.}
\label{fig:angledist}
\end{figure}

Torsional angle distributions show consistent trends that reinforce the above hypothesis.
Flexible chains show sharp peaks at $\psi=0^{\circ}$, $55^{\circ}$, $70^{\circ}$, $110^{\circ}$, $125^{\circ}$, and $180^{\circ}$.
These angles have been shown in previous studies of athermal chains \cite{karayiannis10} to correspond to collapsed, locally polytetrahedral conformations.
As chain stiffness increases, these maxima gradually disappear, and are replaced by a single broad maximum.  
The first maximum occurs at finite $\psi$ for glassforming systems, and at $\psi = 0$ for systems that combine at least some close-packed local with at least intermediate-scale nematic order; cf.\ Figs.\ \ref{fig:nemordervsT}-\ref{fig:Fbb}.
All of these trends are consistent with the vanishing of polytetrahedral order that is expected for semiflexible chains that cannot easily adopt compact conformations.

\subsection{Local and global nematic ordering}

The role of angular stiffness on chain shape is indicated in Figure \ref{fig:nemordervsT}(a), which illustrates the variation of persistence length $l_p$ with $k_b$ as well as its evolution with decreasing $T$.
Systems span the range from the flexible ($l_p \sim r_0$) limit for small $k_b$ to the rodlike ($l_p/a = N-1$) limit for $k_b >\sim 10\epsilon$.
Since the angular potential employed here ($U_b(\theta)$; Eq.\ \ref{eq:Ubend}) is minimized for straight chains and systems remain near thermodynamic equilibrium above $T_s$, stiffer chains clearly uncoil and adopt more-extended configurations as $T$ decreases towards $T_s$.

\begin{figure}[htbp]
\centering
\includegraphics[width=2.4in]{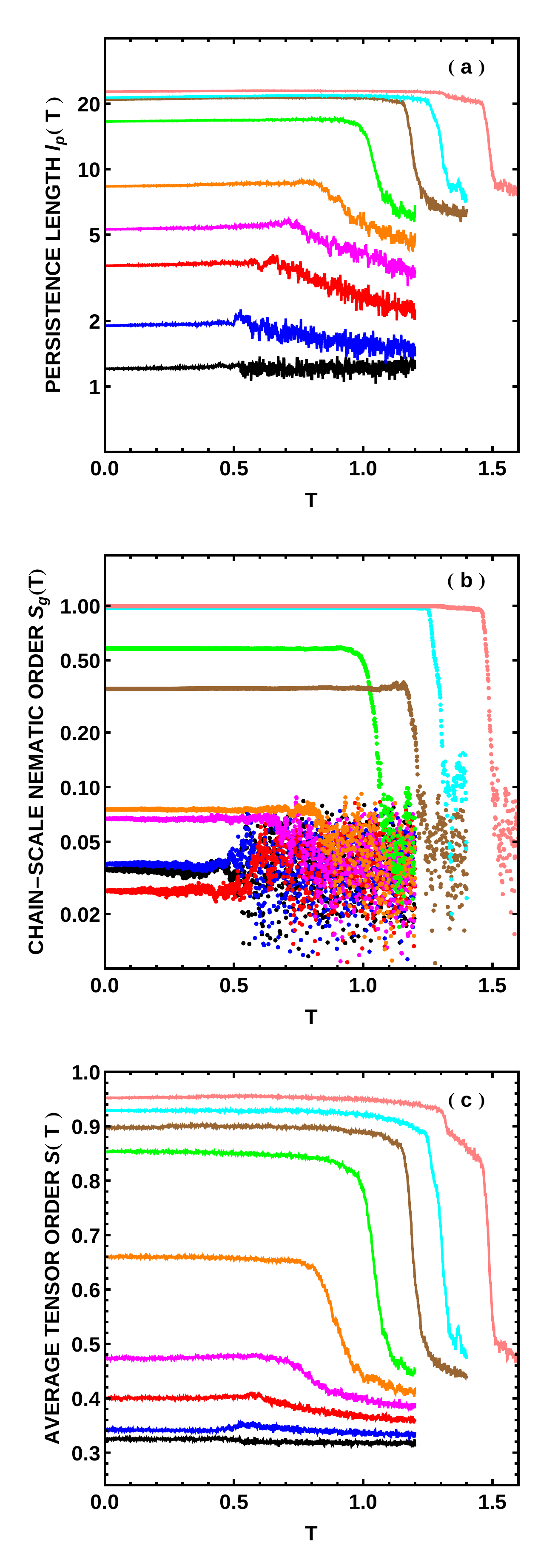}
\vspace{-10pt}
\caption{Thermodynamic signatures of nematic order for selected $N=25$ systems: (a) Persistence length $l_p(T)$ (Eq.\ \ref{eq:perslength}), (b) chain-scale nematic order $S_g(T)$ (Eq.\ \ref{eq:1}), and (c) average tensor order $S(T)$ (Eq.\ \ref{eq:2}).  Colors are the same as in Fig.\ \ref{fig:phi}.}
\label{fig:nemordervsT}
\end{figure}

The same factors that increase $l_p$ also increase chain- and bond-level nematic order.
In Figure \ref{fig:nemordervsT}(b) we present results for the temperature dependence of the chain-level nematic order $S_g(T)$.
For all $T$, $S_g$ increases monotonically with increasing $k_b$ (except for $k_b = 7.0$ systems, which are more ordered than $k_b=8.5$ systems because the multidomain-nematic ordering of the latter reduces $S_g$; see below.)
At high $T > T_s$, ordering oscillates because of the high melt-state mobility of the unentangled chains and the slow cooling rate employed.
As expected, in the flexible limit, chain end-to-end vectors remain randomly oriented for all $T$; no significant ordering at this scale takes place upon crystallization into the RWCP phase.
Glass-forming systems display similar behavior; large-scale chain configurations get ``frozen in'' upon vitrification.
In sharp contrast, stiffer chains display a dramatic increase in $S_g$ upon cooling as chain-scale order transitions from isotropic to nematic.
Indeed, for all but the stiffest chains, the isotropic$\to$nematic transition \textit{drives} crystallization as follows: when chains align, they pack more efficiently, and thus $\phi$ increases.
This densification drives these systems above the characteristic crystallization density $\phi_{cryst}(k_b)$, and crystallization into the NCP phase occurs spontaneously, i.e.\ $T_{cryst} = T_{ni}$ for $7\epsilon <\sim k_b <\sim 10\epsilon$.
Very stiff chains (approaching the rodlike limit) exhibit a separate  isotropic$\to$nematic transition at temperatures above $T_s$.

Figure \ref{fig:nemordervsT}(c) shows the temperature dependence of the average bond-level nematic order $S(T)$.
$S$ increases upon cooling, indicating increasing \textit{local} alignment of chains at the bond scale.
The increases are especially dramatic for NCP-forming systems, but notably, also occur for glass-forming systems (i.e.\ they track the increases in $l_p$), indicating that even the glassforming systems considered here possess a degree of local nematic order, inherited from the melt.
The finite value of $S$ in the flexible limit arises from two factors: (i) topological (chain uncrossability) constraints dictate that nearby bonds can not freely orient with respect to each other, and (ii) in the RWCP crystal, chain segments tend to align preferentially along locally favored directions of their corresponding HCP or FCC crystallites.

\begin{figure}[htbp]
\centering
\includegraphics[width=2.6in]{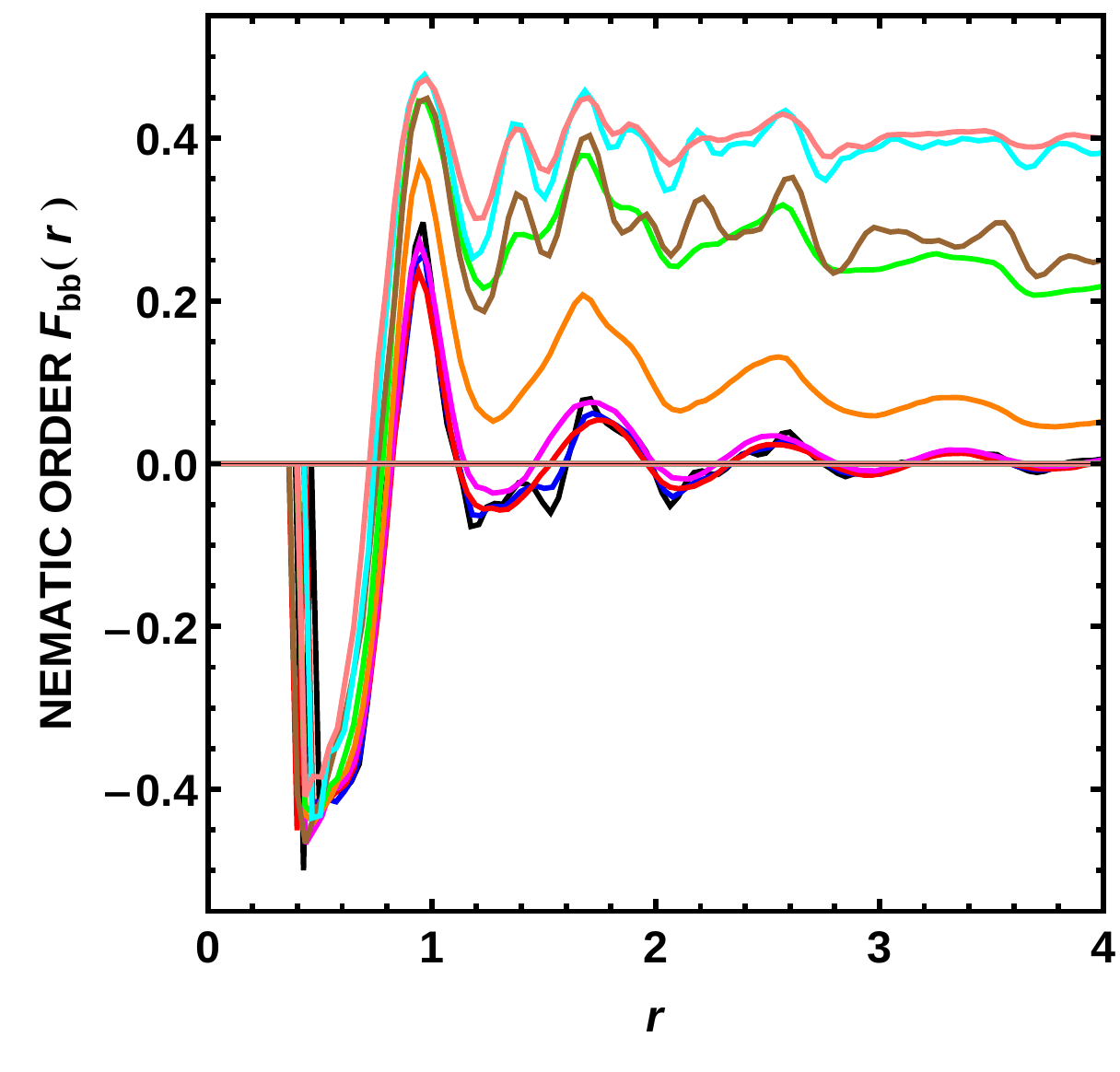}
\vspace{-10pt}
\caption{Nematic order $F_{bb}(r)$ (Eq.\ \ref{eq:bonddircorr}) for selected $N=25$ systems at $T=0$.  Colors are the same as in Fig.\ \ref{fig:phi}.}
\label{fig:Fbb}
\end{figure}

We conclude our discussion of nematic ordering by presenting results for the $k_b$-dependence of the spatial correlation $F_{bb}(r)$ of the bond-vector orientations.  
Results for $T=0$ end states of our cooling runs are shown in Figure \ref{fig:Fbb}.
All $k_b$ display a ``correlation hole'' at small $r$ corresponding to the fact that excluded volume prevents dimer pairs from aligning; the closest allowed separation corresponds to a ``crossed'' configuration which has $F_{bb} = -1/2$.
Similarly, in the densely packed systems considered here, dimer pairs separated by approximately one monomer are preferentially aligned. 
At larger distances, results are highly $k_b$-dependent.
Flexible and glassforming systems exhibit frozen-in, liquid-like order.
Long range chain order sets in for $k_b >\sim 5\epsilon$ and increases rapidly with increasing $k_b$ until (as discussed further below) aligned chains form a single nematic domain at adequately high $k_b$ values.  
Finally, note that the onset of mid-range nematic order at $k_b \sim 6$ corresponds to the crossover from higher to lower values of $\phi_s^{eff}$.

\subsection{(Dis)ordered morphologies formed under cooling; $k_b$-dependence}

\begin{figure*}[htbp]
\centering
\includegraphics[width=6.5 in]{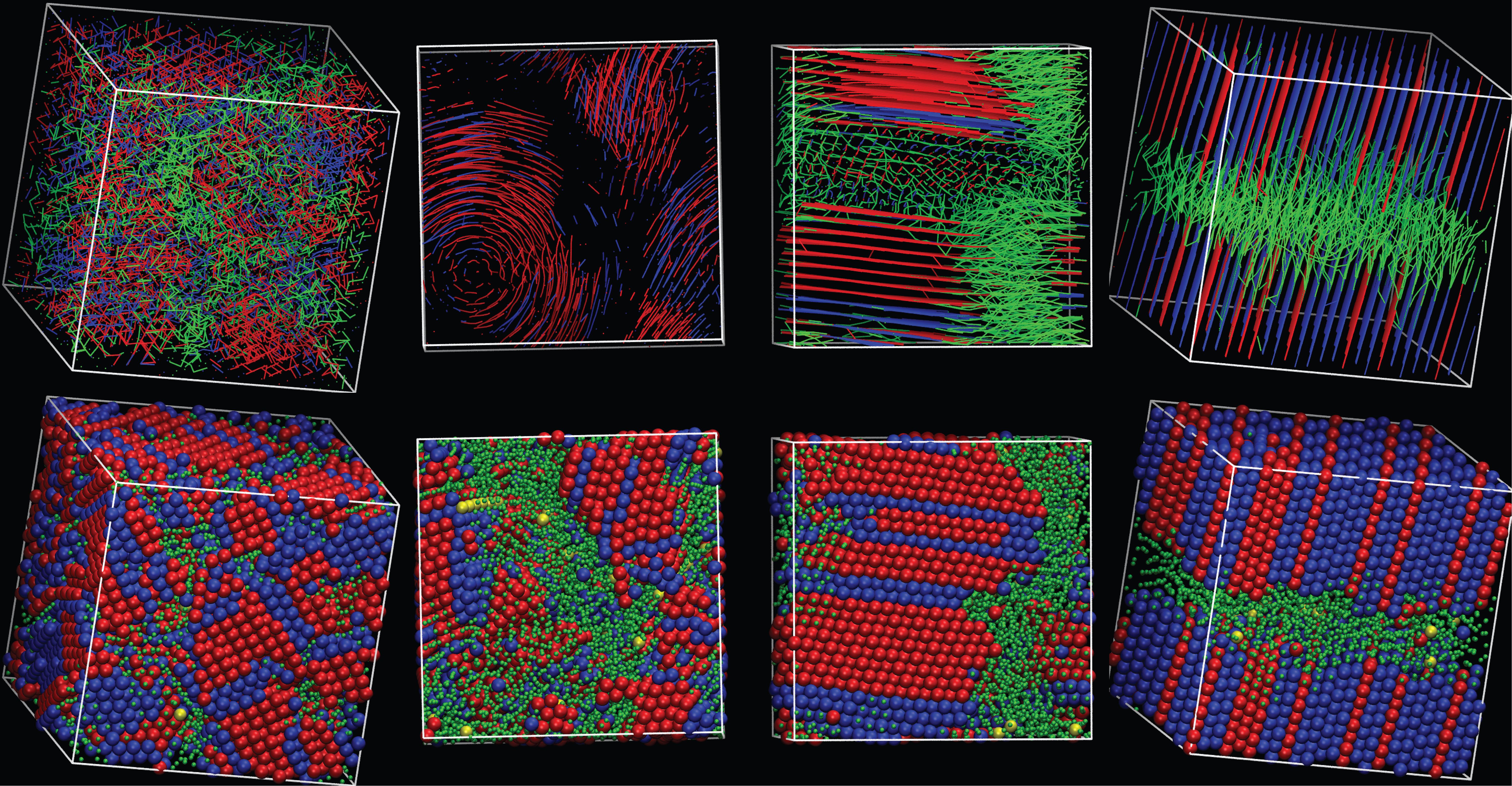}
\caption{Snapshots of the $T=0$ end states of cooling runs for selected $N = 25$ systems. From left to right: $k_{b}/\epsilon = 0.0, 7.0, 8.5\ \rm{and}\ 10.0$. In the upper panel, chains shown as lines, while in the lower panel, monomers are shown as spheres. Chain segments/monomers are color-coded according to the CCE-based norm \cite{cce09}: red, blue, and green respectively correspond to FCC-like, HCP-like, and "other" (non close-packed) local environments. The radii of the "other" monomers (in the sphere representation) are reduced for visualization purposes. Image created with VMD \cite{vmd01}.}
\label{fig:snapshots}
\end{figure*}

Our model exhibits considerably more complexity than might have been surmised, forming a broad array of semi-crystalline morphologies.
Typical system snapshots of final ($T = 0$) configurations for $N = 25$ systems are shown in Figure \ref{fig:snapshots}.
In the flexible limit ($k_b <\sim 1.5\epsilon$), systems freeze into RWCP grains that are randomly shaped and oriented, and are separated by twin defects and/or heavily stack-faulted interphases \cite{hoy13}, similar to results from previous studies of fully flexible athermal \cite{ni13,karayiannis13,karayiannis10} chains.
In the rod-like limit ($k_b >\sim 10\epsilon$), chains tend to form large crystal grains of mixed fcc and hcp character aligned along a single nematic director field, corresponding to close-packed nematic ordering. 
Defects are also present at the employed cooling rate; for example, the $k_b = 10\epsilon$ system possesses an amorphous interphase that is very similar to the amorphous interlamellar domains found in traditional semicrystalline polymers \cite{strobl07}.

In addition to the RWCP and NCP crystals, we also observe more complex forms of long-range order at intermediate $k_b$.
For $k_{b} = 8.5\epsilon$ chains form two distinct close-packed ``grains'' with different nematic orientations, separated by an amorphous grain boundary.
Finally, for $k_{b} = 7\epsilon$, chains form a well-defined spiral morphology.
Remarkably, the monomer-level structure is close-packed (possesses hcp and fcc similarity) even near the core of the spiral, and remains so as the radial distance from the core increases. 
At this $k_b$, the spiral structure forms slightly above $T_s$, freezes in upon solidification, and serves as a nucleus for close-packed crystal growth.
As described in the Appendix, formation of such spirals is quite robust.
This structure is similar to those recently observed in experimental and phase-field theory studies of polymer blends \cite{okabe98,kyu99,xu06}, and illustrates the wide range of ordered morphologies that can be obtained using simple polymer models with variable chain stiffness.

We conclude by presenting a ``phase'' diagram for our model.
Figure \ref{fig:PD} shows both values of $T_s$ and the morphologies formed during solidification, as a function of $k_b$.
Colors and symbol types represent the ordering of the obtained solid phases.
For all chain lengths, $T_s$ drops slightly from its flexible-limit value \cite{hoy13} as a small bending stiffness is added ($k_b <\sim 2\epsilon$), then increases monotonically with increasing $k_b$; it is worth repeating that while this trend is expected for polymers of increasing stiffness \cite{dudowicz05,kunal08,schnell11}, previous studies have not examined models displaying such a broad range of solid-state morphologies.
As described above, systems freeze into RWCP crystals for the smallest $k_b$ and single-domain NCP crystals for the largest $k_b$, while intermediate values of $k_b$ produce either glass-formation or more complex order.

We emphasize, of course, that Fig.\ \ref{fig:PD} is \textit{not} an equilibrium phase diagram, but rather is simply a representation of $k_b$-dependent solidification for the preparation protocol employed here.  
In particular, the results shown in Fig.\ \ref{fig:QRdep} indicate that faster cooling rates will expand the range of $k_b$ over which glasses are formed to both larger and smaller values; lower cooling rates will produce opposite trends.
Lower $|\dot{T}|$ will also naturally extend the range over which single-domain nematic crystals form to lower values of $k_b$.
A detailed examination of the relative thermodynamic stability of these differently ordered phases would be very interesting, but is beyond the scope of the present work \cite{surmise}.

\begin{figure}[htbp]
\centering
\includegraphics[width=2.6in]{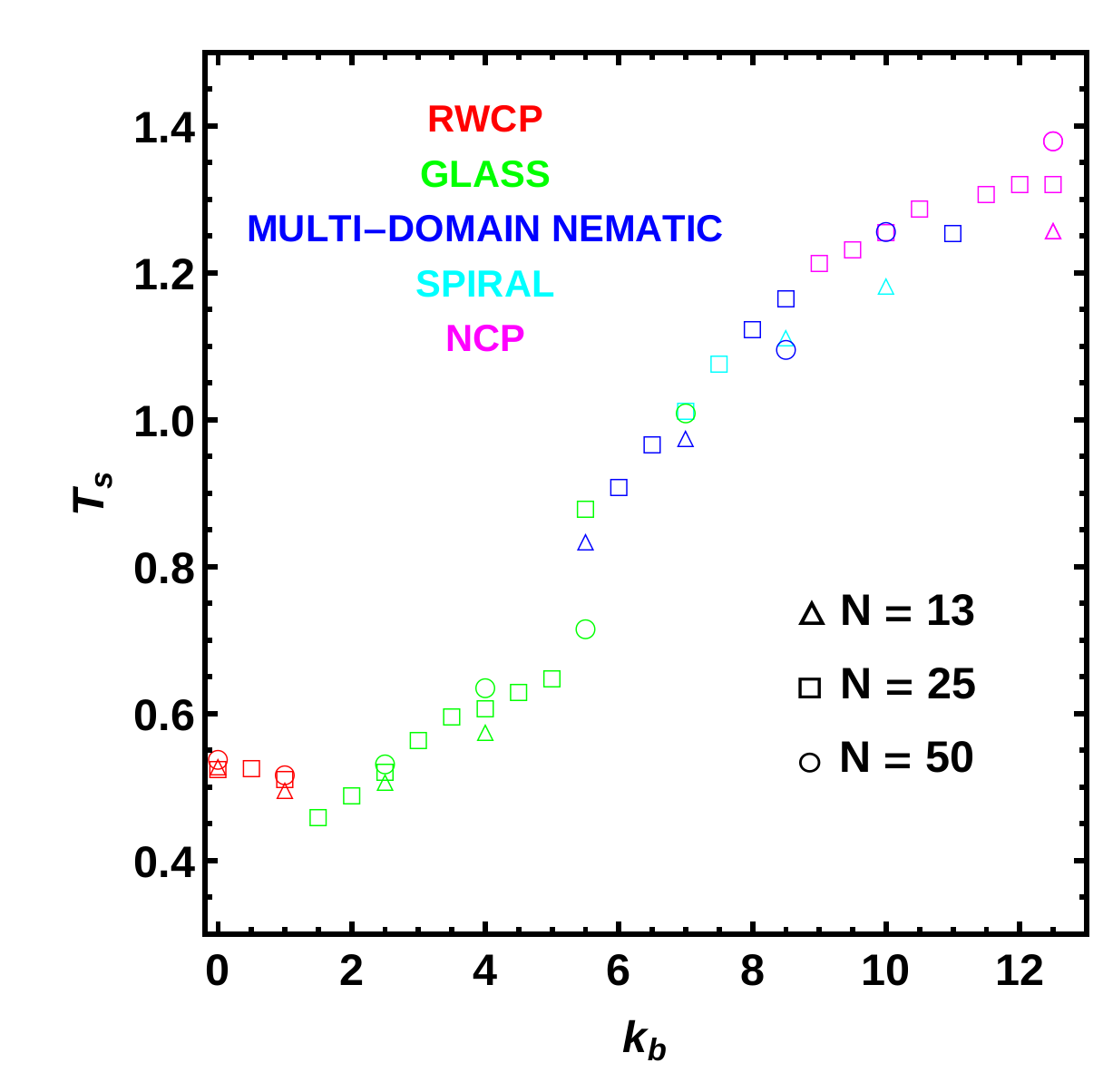}
\caption{``Phase'' or morphology diagram for this model obtained from simulations at low cooling rate ($|\dot{T}|=10^{-6}/\tau$), for chain lengths $N = 13,\ 25,\ \textrm{and}\ 50$.  Multidomain nematic structure is as illustrated in (e.g.) the Fig.\ \ref{fig:snapshots} snapshot for $k_b = 8.5\epsilon$.}
\label{fig:PD}
\end{figure}

\section{Discussion and conclusions}

In this paper, we described the chain-stiffness dependence of the solid-state morphologies formed by model ``soft'' colloidal polymers.
By varying a single interaction parameter ($k_b$), we illustrated dramatic effects of chain stiffness on the competition between crystallization and glass formation.
In fhe flexible-chain (small-$k_b$) limit, monomers occupy the sites of close-packed crystallites while chains retain random-walk-like order.  
At intermediate $k_b$, crystallization is suppressed in favor of glass formation.
As $k_b$ continues to increase, more complex ordered phases such as spirals are also produced, until long-range nematic chain ordering typical of lamellar precursors sets in as the rodlike limit is approached.  
Remarkably, long-range orientational order coexists with close-packing.

The controlling thermodynamics and kinetics of solidification both depend strongly on chain flexibility.
Under cooling at a thermal ``quench'' rate $|\dot{T}|$, relatively flexible chains generally exhibit lower solidification temperatures as well as faster crystallization kinetics (and hence sharper disorder-order transitions) than their stiffer counterparts.  
These dependences, however, are complex and nonmonotonic in $k_b$.
We associated the glass-formation observed at intermediate $k_b$ with the incompatibility of Kuhn-scale structure (i.e.\ bond and torsion angles) with close-packing.
Other factors are probably also highly relevant, including e.g.\ competition between formation of RWCP and NCP crystalline phases.
Future work will consider local structure and dynamics in melts above $T_s$ in order to isolate the microscopic mechanisms underlying this complex, $k_b$-dependent behavior.

In experiments, the attractive interactions between colloidal monomers are typically short-ranged, and weak beyond (at most) $\sim10\%$ of the monomer diameter.
Indeed, most experiments to date on dense colloidal-polymer systems \cite{zou09,brown2012,miskin13,miskin14} have employed ``hard'' (athermal) monomers.
While we have used a Lennard-Jones potential with much longer-ranged attractions, we argue that this difference does not devalue the present work.
Our obtained morphologies agree with those obtained for athermal models in both the flexible \cite{karayiannis09,karayiannis10,karayiannis13,ni13} and rodlike \cite{fynewever98} limits, i.e.\ RWCP and NCP, suggesting that realistic interactions with ranges between the limits of purely repulsive hard spheres and long-ranged-attractive LJ will also produce these morphologies.
It is also well known that many aspects of soft systems can be mapped to hard systems by replacing the ``bare'' packing fraction $\phi$ with an $T$-dependent effective \cite{zhang08} packing fraction $\phi_{eff}(T)$, and vice versa; to aid comparison of our results to systems with different interactions, we reported solidification densities in terms of their effective values $\phi_s^{eff}$.

We therefore claim that the various solid morphologies reported here, and in particular their variation with $k_b$, 
represent a first step towards developing a \textit{qualitative} guide for experiments on real CPs.
For example, our results suggest that by tuning the stiffness and employing a suitable preparation protocol such as ``tapping'' \cite{knight95}, crystalline samples of colloidal or granular polymers may be experimentally realized.
However, it will be critical to investigate the effects upon ordering of factors such as monomer shape (i.e.\ deviations from sphericity) and bond geometry (e.g.\ deviations from $l_0 = r_0$ and different equilibrium bond angles); recent work on both polymeric and nonpolymeric systems suggests that all of these are likely \cite{donev04,damasceno12,ni13} to be very important.
Followup studies aimed at characterizing such effects are in progress \cite{nck15}.

Finally,  we discuss the applicability of the present study to traditional polymers (including familiar synthetic, semicrystalline polymers such as PE and PVA.)
These also crystallize and form hexagonal order in planes perpendicular to the nematic director field in both experiments \cite{strobl07} and simulations \cite{meyer01,meyer02,Vettorel2007}, but do not close-pack.
Preliminary simulations of longer chains have not produced the chain-folded lamellae typical of these systems, perhaps because the lack of torsional interactions removes a necessary thermodynamic driving force; previous work has found that lamella-formation is enhanced by strengthening torsional interactions \cite{lacevic08}.

\begin{appendix}

\begin{figure*}[hbtp]
\includegraphics[height=2in]{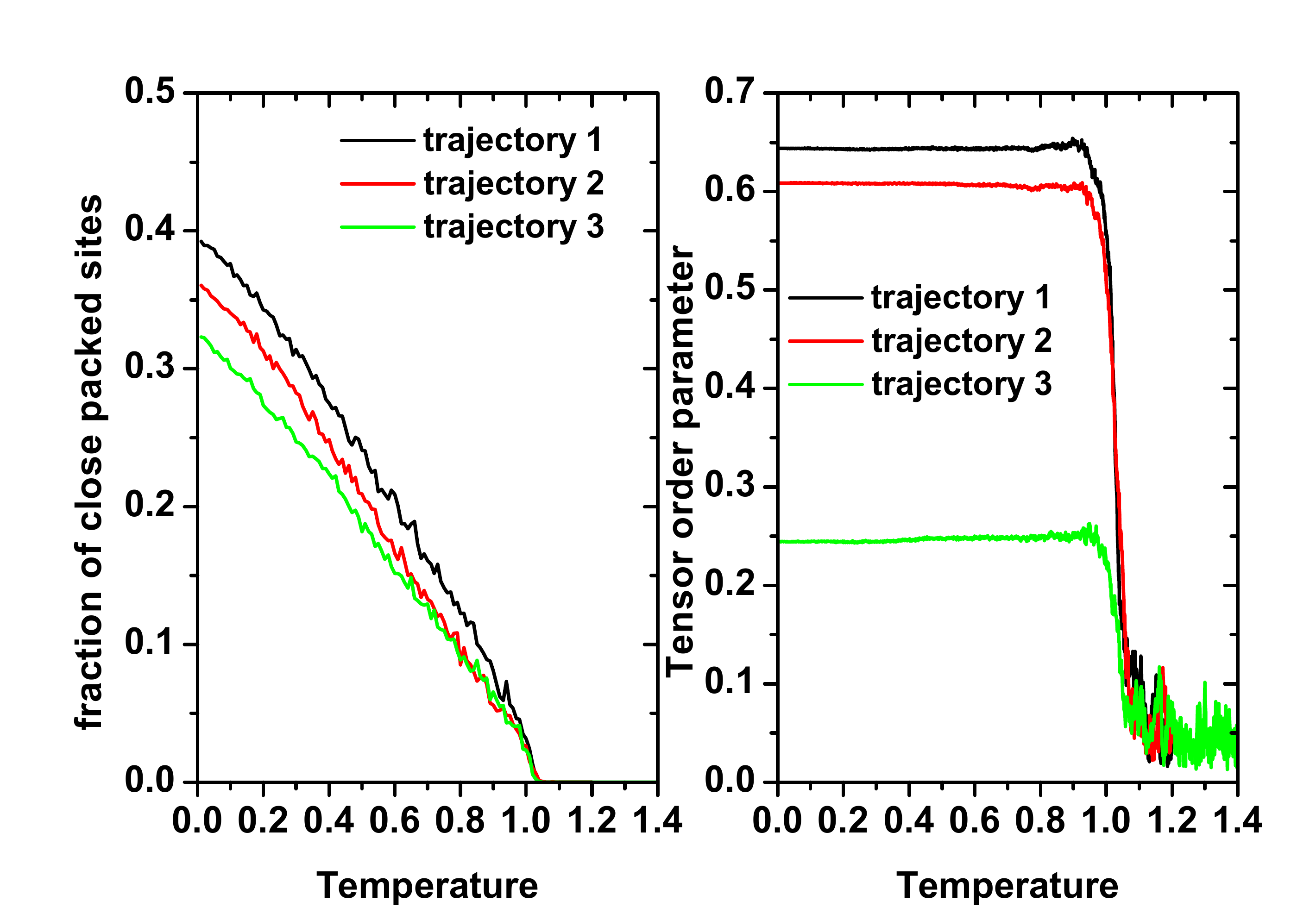}
\hspace{10pt}
\raisebox{30pt}{\includegraphics[height=1.3in]{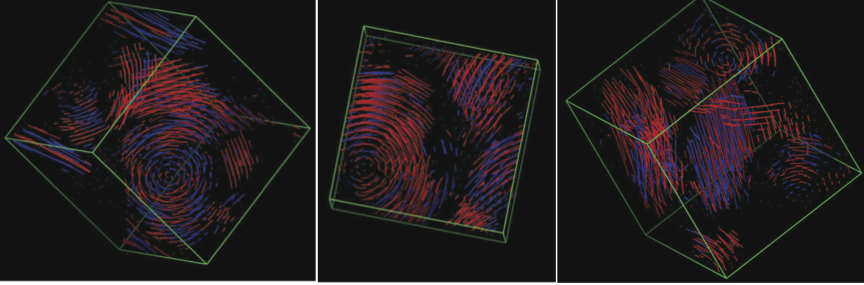}}
\caption{Finite size effects for $N=25$ with $k_b = 7\epsilon$: (Left panels): fraction of close packed sites $f_{cp}$, and global tensor parameter $S_g$ (Eq.\ \ref{eq:1}), versus temperature as obtained from MD runs on three different realizations of $N_{ch} = 500$ systems. Initial configurations were generated using the same protocol. (Right panels): Snapshots of $T = 0$ configurations corresponding to these three realizations. Color coding and the ``line'' representation are the same as Fig. \ref{fig:snapshots}, but for clarity, only close packed (HCP or FCC) chain segments are shown.}
\label{fig:3samples}
\end{figure*}

\section{Finite size effects and spiral morphologies}

Small variations in the initial states of samples may produce larger differences in their final states, especially if the initial states are near a phase transition.
Here we discuss how such variations can affect the morphologies produced, particularly near a phase boundary, focusing on finite-system-size effects.
We consider the sample-to-sample variations of $N=25$, $k_b = 7\epsilon$ systems.
Per Figure \ref{fig:PD}, this value of $k_b$ is in the spiral-forming range, but near the ``transitions'' to other complex ordered phases.
The left panels of Figure \ref{fig:3samples} shows how the fraction of close packed sites $f_{cp}(T)$ and the global tensor parameter $S_g(T)$ (Eq.\ \ref{eq:1}) evolve during cooling for three different (but identically prepared) realizations.
All three trajectories crystallize at practically identical $T = T_{cryst}$ and $\phi=\phi_{cryst}$.
However, the fraction of ordered sites in the final $T=0$ states differs by approximately $15\%$, and the nematic order parameter by a factor $\sim2.5$, between the two extreme cases.
Specifically, very well developed spirals can be observed in the leftmost two ``snapshot'' panels of Fig. \ref{fig:3samples}.
In the rightmost panel, the characteristic ring of the spiral morphology is less developed, and co-exists with nematic domains.
While there are clearly finite-size effects for these systems, these results also make clear that the spiral ``phase'' is quite robust.
Preliminary simulations of much larger systems indicate that the spirals are not artifacts of the periodic boundary conditions (as is sometimes found in systems under severe confinement \cite{pickett00}).
Thus we consider such effects a feature rather than a deficiency of our model since real colloidal polymers are typically out of equilibrium and so are expected to exhibit sample-to-sample variations in typical experiments. 

\end{appendix}


\begin{thebibliography}{67}
\bibitem[{pre()}]{prefac}
\bibinfo{note}{There is some freedom of choice \cite{zhang08} in the prefactor
  of $k_BT$ in Equation \ref{eq:phiseff}. Here we have chosen it to be $1.1$.}
\bibitem[{foo()}]{footSsubcell}
\bibinfo{note}{For the the purpose of calculating $S$, simulation cells are
  divided into $(N_{x} \times N_{y} \times N_{z})$ cubic sub-cells. We chose
  $N_x = N_y = N_z = 14$.}
\bibitem[{sur()}]{surmise}
\bibinfo{note}{Nevertheless, at this point it is worthwhile to compare the
  nonequilibrium results presented in Figure \ref{fig:PD} with those that might
  be ``naively'' expected (from simple equilibrium-thermodynamic arguments) to
  be valid in the limit $|\dot{T}|\to 0$. The $T=0$ ground state for any finite
  $k_b$ is nematic and close-packed (NCP) since ideal NCP packing minimizes
  both bending and pair energies. As temperature increases, entropy favors
  adoption of random-walk-like chain structure, and there will be a
  nematic$\to$isotropic transition at some temperature $T_{ni}(k_b)$ that must
  increase with increasing $k_b$. This transition must cross the
  crystallization line $T_{cryst}(k_b)$ at some characteristic value $k_b^{*}$.
  For $k_b < k_b^*$, the stable crystal should be NCP at low $T$ and RWCP at
  higher $T < T_{cryst}$, while for $k_b > k_b^*$ NCP could plausibly be the
  only stable solid phase. The nucleation barrier for the
  RWCP$\leftrightarrow$NCP solid-solid transition will be extremely high, and
  this transition is not expected to be observed in either simulations or
  experiments. However, for $k_b \simeq k_b^{*}$, it is easy to imagine that
  there is a competition between RWCP and NCP local structure that can suppress
  the formation of ordered crystalline solids.}
\end{thebibliography}

\end{document}